\documentclass[aps,prd,twocolumn,superscriptaddress,amsmath,amssymb,nofootinbib]{revtex4-2}

\usepackage{slashed}
\usepackage{graphicx}
\usepackage{color}
\usepackage[breaklinks,colorlinks=true]{hyperref}
\usepackage{cancel}
\usepackage[normalem]{ulem}
\usepackage{wela}
\usepackage[T1]{fontenc}

\usepackage{fontenc}

\usepackage{bbding}
\usepackage{lettrine}
\usepackage[a-1b]{pdfx}
\usepackage{setspace}
\usepackage{upgreek}

\usepackage{braket}
\usepackage{quoting}
\usepackage[normalem]{ulem}
\usepackage[multiple]{footmisc}
\usepackage{mathtools}
\usepackage{empheq}
\usepackage{hyperref}
\usepackage{xcolor}
\usepackage[toc,page]{appendix}
\usepackage{parcolumns}
\usepackage{mathrsfs}

\newcommand{\E}{\mathcal{E}}

\newcommand{\bea}{\begin{eqnarray}}
\newcommand{\eea}{\end{eqnarray}}
\newcommand{\beq}{\begin{equation}}
\newcommand{\eeq}{\end{equation}}

\newcommand{\bega}{\begin{eqnarray}}
\newcommand{\eega}{\end{eqnarray}}

\newcommand{\sn}{\mathtt{sn}}

\newcommand{\dn}{\mathtt{dn}}

\DeclareMathSymbol{\Epsilon}{\mathalpha}{operators}{"45}
\DeclareMathSymbol{\Kappa}{\mathalpha}{operators}{"4B}

\begin{document}
\title{Quantum vacuum effects in non-relativistic quantum field theory}

\author{Matthew Edmonds}
\email{m.edmonds@uq.edu.au}
\affiliation{ARC Centre of Excellence in Future Low-Energy Electronics Technologies,
School of Mathematics and Physics, University of Queensland, St Lucia, QLD 4072, Australia}
\affiliation{Research and Education Center for Natural Sciences,
  Keio University, 4-1-1 Hiyoshi, Yokohama, Kanagawa 223-8521, Japan}

\author{Antonino Flachi}
\email{flachi@phys-h.keio.ac.jp}
\affiliation{Department of Physics,
  Keio University, 4-1-1 Hiyoshi, Yokohama, Kanagawa 223-8521, Japan}
\affiliation{Research and Education Center for Natural Sciences,
  Keio University, 4-1-1 Hiyoshi, Yokohama, Kanagawa 223-8521, Japan}

\author{Marco Pasini}
\email{marco.pasini@uniud.it}
\affiliation{Department of Mathematics, Computer Science, and Physics, University of Udine, Via delle Scienze 206
I-33100 Udine, Italy}
\affiliation{Research and Education Center for Natural Sciences,  Keio University, 4-1-1 Hiyoshi, Yokohama, Kanagawa 223-8521, Japan}

\begin{abstract}
Nonlinearities in the dispersion relations associated {with} different interactions designs, boundary conditions and the existence of a physical cut-off scale can alter the quantum vacuum energy of a nonrelativistic system nontrivially. As a material realization of this, we consider a 1D-periodic rotating, interacting non-relativistic setup. The quantum vacuum energy of such a system is expected to comprise two contributions: a fluctuation-induced 
quantum contribution and a repulsive \textit{centrifugal-like} term. We analyze the problem in detail within a complex Sch\"odinger quantum field theory with a quartic interaction potential and perform the calculations non-perturbatively in the interaction strength by exploiting the nonlinear structure of the associated nonlinear Schr\"odinger equation. Calculations are done in both zeta-regularization,  
as well as by introducing a cut-off scale. We find a generic, regularization-independent behavior, where the competition between the interaction and rotation can be balanced at some critical ring-size, where the quantum vacuum energy has a maxima and the force changes sign. The inclusion of a cut-off smoothes out the vacuum energy at small distance but leaves unaltered the long distance behavior. We discuss how this behavior can be tested with ultracold-atoms. 
\end{abstract}
\maketitle

\section{Introduction} 

In quantum field theory, the canonical quantization scheme does not fix the order of non-commuting operators in the Hamiltonian, leaving a residual divergent ``\textit{zero-point energy}'' contribution to the energy density (in natural units):
\bea
\mathscr{E} = {1\over 2} \sum_n \omega_n,
\label{eq:1}
\eea
with $\omega_n$ representing the frequencies of the quantum fluctuations. Wick's normal ordering is then used to enforce a specific order of operators' products, resulting in the subtraction of this infinite shift from the vacuum expectation value (vev) of the Hamiltonian that will then vanish. This has the consequence that the quantum vacuum, \textit{so defined}, does not carry energy, linear or angular momentum. Such a procedure is usually justified by saying that a constant shift in the energy cannot be measured, although this view is not entirely tenable as any finite  energy is, in principle, measurable due to its gravitational effect. In relativistic quantum field theory, a better justification follows from the fact that the expectation value of the Hamiltonian in the \textit{noninteracting} vacuum (i.e., in absence of external fields or interactions) must vanish for the Hamiltonian, \textit{a} generator of the Poincar\'e group, to satisfy the correct commutation rules. Then, the usual notion of a noninteracting vacuum as a state devoid of energy follows, justifying the use of normal ordering \cite{Takahashi:1969,Plunien:1986}.

Even without calling gravity into question \cite{Ford:1975}, a variety of quantum vacuum phenomena, most notably the Casimir effect \cite{Casimir:1948}, clearly demonstrates some level of inadequacy of the above definition of an \textit{empty} physical vacuum \textit{tout court}. In the original version of the Casimir effect, for example, this was evident owing to the imposition of boundary conditions on the quantum fluctuations of the electromagnetic field in the presence of perfectly conducting, parallel plates, resulting in 
an attractive force between the plates. More general (and realistic) si\-tua\-tions are not different, as boundary conditions result from quantum fields existing in interaction with other fields, and modify the spectrum of the quantum fluctuations, thus changing the zero-point energy. 

These arguments converge into Casimir's definition of the energy of the quantum va\-cuum ${E}_{vac}$ as the difference between the zero-point energies in the presence, $E\left[{\mathcal \partial B}\right]$, and in absence, $E\left[{\emptyset}\right]$, of boundaries:
\bea
{E}_{vac} = E\left[{\mathcal \partial B}\right] - E\left[{\emptyset}\right]. 
\label{eq:2}
\eea
Such a definition is compatible with the vanishing of the vev of the Hamiltonian in the noninteracting vacuum (i.e., no boundary) and gives a calculable recipe (within any regularization scheme) of the quantum vacuum energy in response to changes in external conditions \cite{Plunien:1986,Milton:2001,Bordag:2009}. This 
view on the complexity of the vacuum has been vindicated during the past quarter of a century by many successful experiments starting with \cite{Lamoreaux:1997,Mohideen:1998} (see also Ref.~\cite{Gong:2021} for a recent additional list of examples of applications to nanophotonics, nanomechanics, and chemistry).

A less explored question concerns the quantum va\-cuum energy in non-relativistic systems (see, for some discussions, Refs.\cite{Toms:2012,Toms:2002,Nakayama:2023,Cougo-Pinto:2002,Corradini:2021yha,Kolomeisky:2013zra}). The answer might {seem} simple, since in a non-relativistic context there is no issue associated with antiparticles or the ordering of the operators, suggesting that the zero-point energy can be safely ignored. However, this is not the case in general. Even from the vantage point of the original Casimir effect, the story remains subtle because the quantum vacuum energy emerges from deformations of the electromagnetic quantum fluctuations, and no simple non-relativistic limit can be taken: the photon is massless and propagates at the speed of light. 

However, in a non-relativistic set-up one can imagine emergent degrees of freedom, constrained by boundaries, and how these could give rise to non-trivial quantum vacuum phenomena and a number of works have explored such question, particularly in the context of quantum liquids and Bose-Einstein condensation where (\ref{eq:1}) contributes to the zero temperature thermodynamic potential (on top of the classical ground state contribution); see, for example, Refs.~\cite{Recati:2005,Roberts:2005,Edery:2006,Schiefele:2009,Biswas:2010,Schecter:2014,Schecter:2015,VanThu:2017,Marino:2017,Song:2021,Song:2022}. 

There are at least two reasons why in a non-relativistic setting the situation is far from obvious.  The first is that any time we are in the presence of interactions and non-trivial boundary conditions, the frequencies $\omega_n$ in (\ref{eq:1}) develop a non-trivial dependence on the ground state of the system. 
This can be seen using  the background field method (see Ref.~\cite{Toms:2002}), although computing the frequencies within this framework becomes a hard task. Earlier calculations relying on a perturbative expansion around small coupling exist \cite{Ford:1979b,Toms:1980a,Peterson:1982} and more recently Ref.~\cite{Bordag:2021} has developed a way to compute the quantum vacuum energy for a relativstic $1+1$ dimensional scalar field theory without relying on expansions in powers of the interaction strength (see also Refs.~\cite{Flachi:2013b,Flachi:2017b,Flachi:2021b,Flachi:2023prd}). The second reason has to do with the regularization. In the relativistic case, the quantum vacuum energy emerges from the summation of the entire spectrum as in (\ref{eq:1}); this summation is divergent and must be regularized. A subtlety with this is due to the existence of a physical cut-off that may alter the spectral sum in (\ref{eq:1}). Within a lattice approach this should be possible (see Ref.~\cite{Nakayama:2023}), however it is not at all obvious how to do this within an effective field theory approach. It is certainly an interesting question to ask whether any remnant of the quantum vacuum energy remains in the non-relativistic limit.  

When the model under consideration is nonlinear, the difference in the dispersion relation due to the presence of interactions, the presence of external forcing (e.g., rotation), a physical cut-off scale and boundary conditions are all factors that together conjure to induce intricate behaviors in the quantum vacuum energy. Here we look at the above questions within the paradigmatic nonlinear Schr\"odinger equation. Our approach to compute the quantum vacuum energy 
exploits the integrability structure of the nonlinear Schr\"odinger equation associated with our problem. The calculations are done both using zeta-regularization including the contribution from the whole spectrum, as well as a more physical regularization scheme where the spectral sums are modulated by a frequency dependent window-function that suppresses the contribution of the high-energy modes, leaving a dependence on a physical cross-over scale. As we shall see the two methods lead to compatible results, with the only expected consequence of the cut-off being that of regularizing the vacuum energy at short distance. In conclusion, we will describe how our predictions can, in principle, be measured experimentally with cold-atom rings.

\section{Non-relativistic Schr\"odinger model} 

We shall consider a system of non-relativistic interacting bosons, described by a complex Schr\"odinger quantum field $\Phi ={(\phi_1+ i\phi_2)}/{\sqrt{2}}$, with $\phi_1,\phi_2 \in \mathbb{R}$, confined to a 1D ring of radius $R$, rotating with constant angular velocity $\Omega$. We assume that the periodicity of the ring is externally broken by the presence of a barrier that we describe by imposing Dirichlet boundary conditions at one point on the ring. The Lagrangian density is 
\begin{equation}
\begin{split}
\label{eq:4}
\mathcal{L} &=\frac{i}{2}\left(\Phi^\dagger \dot{\Phi} -\Phi \dot{\Phi}^\dagger\right)+
\frac{i}{2}\Omega\left(\Phi^\dagger \Phi^\prime -\Phi \Phi^{\dagger\prime}\right)- \\
&\frac{1}{2mR^2}\Phi^{\dagger\prime}\Phi^\prime - \frac{\lambda}{4}\left(\Phi^\dagger\Phi\right)^2,
\end{split}
\end{equation}
where $0 \leq \varphi \leq 2\pi$, $x=R \varphi$, $\dot{ } = d/dt$ and ${}^\prime = d/d\varphi$. We adopt units of $\hbar =1$. Expression (\ref{eq:4}) represents the Lagrangian density of an observer co-rotating with the ring. In this reference frame, boundary conditions for the co-rotating observer are time-independent \cite{Chernodub:2012em,Schaden:2012,Ambrus:2014}. 
The following nonlinear Schr\"odinger equation can be derived from (\ref{eq:4}): 
\begin{equation}
\begin{split}
i\dot{\Phi}=-i\Omega\Phi^\prime-\frac{1}{2mR^2}\Phi^{\prime \prime}+\frac{\lambda}{2}\left|\Phi\right|^2\Phi.
\end{split}
\label{eq:7}
\end{equation}
The normal mode decomposition can be carried out by looking for stationary solutions of the form 
\begin{equation}
\begin{split}
\Phi\left(t,\varphi\right)= e^{-i\omega_p t}f_p\left(\varphi\right).
\end{split}
\end{equation}
This allows us to write the original equation (\ref{eq:7}) as
\begin{equation}
\begin{split}
\label{eq:9a}
0 = \frac{1}{2mR^2}f_p^{\prime \prime}+i\Omega f_p^\prime -\left(\frac{\lambda}{2}\left|f_p\right|^2-\omega_p\right)f_p.
\end{split}
\end{equation}
To solve (\ref{eq:9a}) we decompose $f_p$ as 
\begin{equation}
\begin{split}
\label{eq:9}
f_p\left(\varphi\right)=\rho\left(\varphi\right)e^{i\alpha\left(\varphi\right)}, \,\,\,\,\,\,\,\,\,\,\,\, \text{with} \,\,\rho\left(\varphi\right), \alpha\left(\varphi\right) \in \mathbb{R},
\end{split}
\end{equation}
that leads to
\bea
\hskip -1cm 0&=& \frac{\rho^{\prime\prime}}{2mR^2}+\left(\omega_p-\frac{\alpha^{\prime \, 2}}{2mR^2}-\Omega\alpha^\prime\right)\rho-\frac{\lambda}{2}\rho^3, ~~~
\label{eq:10}
\\
\hskip -1cm 0&=& \frac{\alpha^{\prime\prime}\rho}{2mR^2}+\frac{1}{mR^2}\alpha^\prime\rho^\prime+\Omega\rho^\prime.
\label{eq:11}
\eea
The above system of equations can be solved analytically, first obtaining $\alpha'$ in terms of $\rho$ from Eq.~(\ref{eq:11}),
\bea
\alpha^\prime=\beta=\frac{C}{\rho^2}-mR^2\Omega,
\eea
($C$ is an integration constant) and then substituting $\alpha^\prime$ in Eq.~(\ref{eq:10}); this gives rise to a cubic nonlinear equation in $\rho$ that can be solved in terms of Jacobi elliptic functions. Imposition of the boundary conditions selects the solution as a Jacobi $\sn$ function and leads to the following quantization conditions for the eigenfrequencies. The procedure is straightforward but lengthy. For completeness we give all the details in the Appendix and refer the reader to Refs.~\cite{Lakshmanan:2003,Carr:2000::1,Carr:2000::2,Sacchetti:2020,Cominotti:2014} for further details on elliptic equations. The solution can be written as
\begin{equation}
\begin{split}
\label{eq:12}
\Phi\left(t,\varphi\right)=A_n\, e^{-i\omega_n t}e^{-i\left(mR^2\Omega\varphi-\pi/4\right)}\sn\left(q_n\varphi,k_n\right),
\end{split}
\end{equation}
with the normalization factor $A_n$ expressed in terms of elliptic integrals of first and second kind, $K(z)$ and $E(z)$ respectively, 
\bea
A_n^2=\frac{k_n^2}{2\pi R\left(1-{E\left(k_n\right)}/{K\left(k_n\right)}\right)}.
\label{eq:13}
\eea
The momentum $q_n$ and the elliptic modulus $k_n$ are quantized according to the following relations
\bea
\label{eq:14}
q_n&=&\frac{n}{\pi}K\left(k_n\right), ~~~ n \in \mathbb{N},\\
\lambda m R\frac{\pi}{4n^2} &=& K\left(k_n\right)\left(K\left(k_n\right)-E\left(k_n\right)\right),
\label{eq:15}
\eea
where (\ref{eq:14}) comes from the periodicity of the solution and (\ref{eq:15}) is derived from the first integral of the equation of motion. Finally, the eigenfrequencies are given by
\bea
\omega_n = \left(1+k_n^2\right)q_n^2/(2mR^2) - {m}R^2\Omega^2/2. 
\label{eq:16}
\eea
Details on how to derive Eqs.~(\ref{eq:12}), (\ref{eq:13}), (\ref{eq:14}), (\ref{eq:15}) and (\ref{eq:16}) are given in the Appendix.

\section{Quantum vacuum energy and spectral asymptotics} 
\begin{figure}[h]
\centering
\includegraphics[width=1\columnwidth]{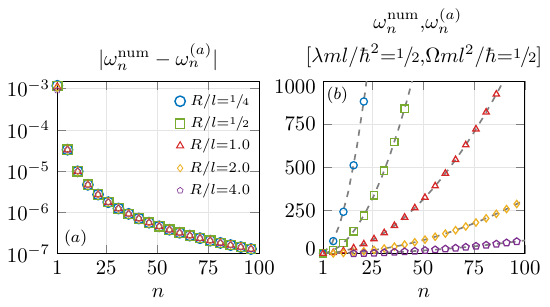}
\caption{(color online) Comparison between (\ref{eq:16}) and the eigenvalues computed numerically. Panel (a) shows the absolute difference between the asymptotic and exact nonlinear eigenvalues, while (b) shows individual datasets for fixed $R$. Coloured data in (b) were obtained numerically, while the grey dashed lines were calculated from Eq.\eqref{eq:16}. Here $\Omega Rml/\hbar = 0.5$ and $\lambda = ml/\hbar^2 = 0.5$ throughout.}
\label{Fig1}
\end{figure}
In the following, we illustrate how to compute the quantity (\ref{eq:1}) for the present case. A non-renormalized expression for the quantum vacuum energy can be written as follows (see Ref.~\cite{Toms:2012,odints:book}):
\bea
\mathscr{E}_r(s) = {\mu^s \over 2} \sum_n \omega_n^{1-s},
\label{eq:22}
\eea
where $s \in \mathbb{C}$ is a complex-valued regularization parameter and $\mu$ is a renormalization scale with dimension of energy. The index $r$ is a reminder that (\ref{eq:22}) refers to the co-rotating frame. The eigenvalues $\omega_n$ are given in terms of the nonlinear, coupled algebraic equations (\ref{eq:14}), (\ref{eq:15}) and (\ref{eq:16}). The regularization of (\ref{eq:22}) is done by finding a representation that converges in some region of the complex-$s$ plane, followed by analytical continuation to the physical value $s\to 0$. Here, we use the spectral asymptotics of the eigenvalues and express (\ref{eq:22}) as  
\bea
\mathscr{E}_r(s) = \Delta + \tilde{\mathscr{E}_{r}}(s),
\label{eq:23}
\eea
where
\bea
\tilde{\mathscr{E}_{r}}(s) =  {\mu^s \over 2} \sum_n \left(\omega^{(a)}_n\right)^{1-s},
\label{eq:24}
\eea
and
\bea
\Delta = {1\over 2} \sum_n \left(\omega_n - \omega^{(a)}_n\right).
\label{eq:25}
\eea
The quantity $\omega^{(a)}_n$ represents the asymptotic expansion of the eigenvalues $\omega_n$ as a function of the quantum number $n$. If the asymptotic expansion includes all terms up to $O(1/n^2)$ as we shall do here, then $\Delta \sim O(1/n^2)$, i.e. (\ref{eq:25}), and thus converges for $s\to 0$ (in formula (\ref{eq:25}) we have already set $s\to 0$). Such a procedure simply confines the divergences to $\tilde{\mathscr{E}_{r}}(s)$ that will need explicit regularization.
The first step of the process is to obtain the asymptotic behavior of the eigenvalues. This can be obtained numerically, but it is not difficult to find its analytical form. Since the left hand side of Eq.(\ref{eq:15}) converges to zero for $n\to \infty$, while the right hand side, as a function of $k_n$, goes to zero only in the limit $k_n \to 0$, while decreasing monotonically for increasing $k_n>0$, the right hand side of (\ref{eq:15}) for small $k_n$ gives the relevant limit to capture the large $n$ asymptotic behavior $k_n^2 \approx {2 \lambda m R / (\pi n^2)}$. This result used in conjunction with (\ref{eq:14}) and (\ref{eq:16}) allows {us} to readily extract the leading asymptotic behavior of $\omega_n$:
\bea
\omega^{(a)}_n = n^2/\eta^2 + \rho^2 + O\left(1/n^2\right),
\label{eq:27}
\eea
where $\eta^2 = {8 m R^2}$ and $\rho^2 = {3 \lambda / (8 \pi R)} - {m R^2 \Omega^2 / 2}$. 
Fig.\ref{Fig1} shows a comparison between the eigenvalues computed numerically and their asymptotic counterpart. The large-$n$ scaling of the eigenvalues is consistent with Weyl’s law that in the present case predicts a leading large-$n$ behavior of the $\omega_n$ scaling as $n^2$ and independent of $\lambda$ \cite{Baltes-Hilf:1976}. 
Using (\ref{eq:27}),
\bea
\tilde{\mathscr{E}_{r}}(s) =  {\left({\mu\eta^2}\right)^s\over 2\eta^2} \sum_n \left( n^2 + \eta^2\rho^2 \right)^{1-s},
\label{eq:30}
\eea
and the Chowla-Selberg representation (See Refs.~\cite{Flachi:2008,Elizalde:cs_rep}),
\bea
\begin{split}
\label{eq:31}
&\sum_{n=1}^\infty \left(n^2 + \gamma^2\right)^{-z}
=-{\gamma^{-2z} \over 2} + {\sqrt{\pi} \over 2} {\Gamma(z-1/2)\over \Gamma(z)} \gamma^{1-2z}\\
&+{2\pi^{z}\over \Gamma(z)}\gamma^{-z+1/2} \sum_{p=1}^\infty 
p^{z-1/2} K_{z-1/2} \left(2 \pi p \gamma\right),
\end{split}
\eea
from which the 
limit $s \to 0$ can be taken to arrive at the following regularized expression
\begin{figure}[t]
\begin{center}
\includegraphics[width=1.0\columnwidth]{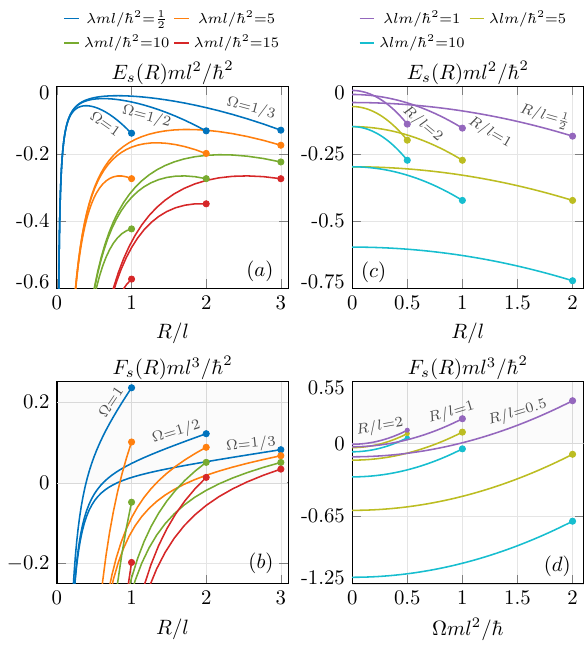}
\end{center}
{\caption{\label{fig:ef} (color online) Quantum vacuum energy and force. (a) and (c) show the quantum vacuum energy, Eq.~\eqref{eq:36} for the same $\lambda$ (colour groups) and varying $\Omega$ (panel a) or $R$ (panel c) (see text labels). (b) and (d) show the corresponding force, Eq.~\eqref{eq:37} for each data set. The light grey shading indicates the parameter regions where the force changes sign from attractive to repulsive. Throughout the paper, the quantity $l$ represents a generic unit-scale.}}
\end{figure}
\bea
\tilde{\mathscr{E}_{r}} =\lim_{s\to 0}\tilde{\mathscr{E}_{r}}(s) =
-{1\over 4} \left({3 \lambda \over 8 \pi R} - {m R^2 \Omega^2 \over 2}\right).
\label{eq:32}
\eea
Thus, the total quantum vacuum energy in the co-rotating frame is given by $E_r = \tilde{\mathscr{E}_{r}} + \Delta$. To get the energy in the laboratory frame $E_s$ one can use $E_s - E_r = \bf \Omega L$, where $L =  - {\partial E_r / \partial \Omega}$ 
is the angular momentum \cite{Flachi:2023prd,Landau:1980,Chernodub:2012em,Schaden:2012}:
\bea
E_s = \Delta - \Omega {\partial \Delta \over \partial \Omega} -{3 \lambda \over 32 \pi R} - {m R^2 \Omega^2 \over 8}.
\label{eq:36}
\eea
The resulting force $F_s = -{\partial E_s / \partial R}$ is
\bea
F_s =  -{\partial \Delta \over \partial R} + \Omega {\partial^2 \Delta \over \partial \Omega \partial R}
-{3 \lambda \over 32 \pi R^2} + {m R \Omega^2 \over 4}.
\label{eq:37}
\eea
Ignoring for the time being the contributions from $\Delta$, the above expression comprises a contribution proportional to $-\lambda/R$ that vanishes for $\lambda \to 0$ and scales as the inverse of the ring size: this is an attractive ``Casimir-like'' contribution. The other contribution {$E_{\mathcal{I}}=\tfrac{1}{2}\mathcal{I}_{\mathcal{R}}\Omega^2$} is proportional to the {moment of inertia of the ring $\mathcal{I}_{R}=mR^2$ with radius $R$}. The vanishing behavior for $\lambda \to 0$ and $\Omega \to 0$ is consistent with the fact that the quantum vacuum energy should vanishes in {the} absence {of} interactions and boundary condtions. The angular velocity appears as the square of $\Omega$, and this is again consistent with the fact that our model does not include parity breaking terms, thus the energy should be symmetric wrt $\Omega \leftrightarrow -\Omega$. 
The force vanishes at the critical radius 
\bea
R_{crit} \approx \sqrt[3]{3 \lambda \over 8 \pi m \Omega^2},
\label{eq:38}
\eea
with its sign changing from negative-attractive for $R<R_{crit}$ to positive-repulsive for $R>R_{crit}$.
Interestingly, also the way the force scales with the ring size changes with the angular velocity: it scales linearly in the regime of fast rotation, while it scales as the inverse square of the ring size for slow rotation. The symbol ``$\approx$'' in (\ref{eq:38}) indicates that the contribution of $\Delta$ has been ignored. Units of $\hbar$ are restored in the numerics. Fig.~\ref{fig:ef} shows the quantum vacuum energy (panels a and c) as a function of radius $R$ and rotation strength $\Omega$ respectively, while the two lower panels (b and d) show the corresponding force associated with each dataset from (a) and (c). The grey shaded region shows the parameter regime where  
the force is repulsive. 

\begin{figure}[t]
\centering
\includegraphics[width=1\columnwidth]{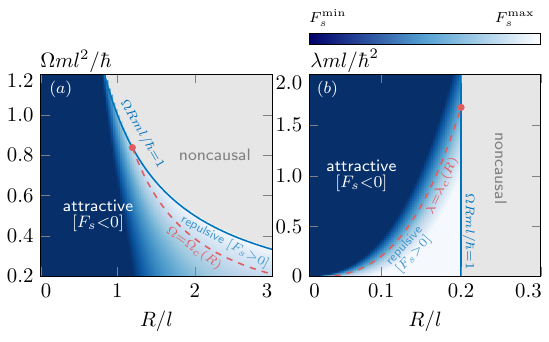}
{\caption{\label{fig:hmap} (color online) Quantum fluctuation induced-force heatmaps. (a) shows Eq.~\eqref{eq:37} in the limit $\Delta=0$ in the $(\Omega,R)$ parameter space for fixed $\lambda ml/\hbar^2=10$. The solid blue line indicates the border between the repulsive and noncausal regions, while the dashed red line indicates the point at which the force changes sign. (b) shows the magnitude of the force in the $(\lambda,R)$ parameter space for fixed $\Omega ml^2/\hbar=5$.}}
\end{figure}

Fig.~\ref{fig:hmap} shows heatmaps of Eq.~\eqref{eq:38} in the {$(R,\Omega)$} and {$(R,\lambda)$} parameter spaces, (a) and (b) respectively. In panel (a) the interaction strength is $\lambda ml/\hbar^2=10$ while the rotation strength is $\Omega ml^2/\hbar=5$ in (b). The solid blue lines in both panels show the border between the repulsive regime and the causality limit defined by $\Omega R ml/\hbar=1$. The red dashed line indicates where the 
force changes sign, obtained from Eq.~\eqref{eq:38}. The red data point in each panel corresponds to the point $(R_{\circ},\Omega_{\circ})$ {in} (a) and $(R_{\circ},\lambda_{\circ})$ {in} (b) where Eq.~\eqref{eq:38} and the causality limit coincide, and 
\begin{equation}\label{eqn:rz}
\big(R_{\circ},\Omega_{\circ}\big)=\bigg(\frac{3}{8\pi}\frac{\lambda ml^2}{\hbar^2},\frac{8\pi}{3}\frac{\hbar^2}{m^2l^3}\frac{1}{\lambda}\bigg).
\end{equation}
The point defined by Eq.~\eqref{eqn:rz} in Fig.~\ref{fig:hmap} shows the maximum rotation strength where repulsive solutions are obtained, then the model of Eq.~\eqref{eq:7} is expected to support a causal repulsive force in the region $\Omega_c<\Omega<R^{-1}(ml/\hbar)$ and $R>R_{\circ}$. Likewise for panel (b) the causal repulsive regime is defined between $0<\lambda<\lambda_{\rm c}$ and $0<R<R_{\circ}$. An analysis, qualitatively similar to Fig.~\ref{fig:hmap}(b), can be done for the {$(R,\Omega)$} parameter space with constant $\lambda$.

A subtle point has to do with how the above results will change in the presence of a cut-off scale associated with a minimal length scale (e.g., the inter-atomic separation scale). 
We address this question by modifying the regularization procedure to include a frequency dependent \textit{window function}. This is implemented by defining 
\bea
\tilde{\mathscr{E}_r} =
{1 \over 2} \sum_n \omega^{(a)}_n \sigma_n(\ell_c),
\label{eq:40}
\eea
and the residual $\Delta$ as
\bea
\Delta = {1\over 2} \sum_n \left(\omega_n - \omega^{(a)}_n\right) \sigma_n(\ell_c).
\label{eq:41}
\eea
Here, we choose the {window function} as follows:
\bea
\sigma_n(\ell_c) = \exp\left({-{\ell_c n^2/ (8 m R^2)}}\right),
\label{eq:42}
\eea
with the argument of the exponential set by the leading large-$n$ asymptotics of the spectrum. The cut-off scale $\ell_c$ determines how high-frequency modes are suppressed. The limit $\ell_c \to 0$ of (\ref{eq:40}-\ref{eq:41}) returns the non-regularized expression for $\tilde{\mathscr{E}_r}$ discussed earlier. While the choice of $\sigma_n(\ell_c)$ is arbitrary, (\ref{eq:42}) allows to write (\ref{eq:40}) as
\bea
\begin{split}
\tilde{\mathscr{E}_r} &=&
-{\rho^2\over 4} + {\rho^2\over 4} \theta_3\left({\ell_c\over \pi \eta^2} \right)
- {1\over 4\pi \eta^2} \theta_3'\left({\ell_c\over \pi \eta^2} \right).\label{eq:44}
\end{split}
\label{eq:31er}
\eea
where $\theta_3$ is the following Jacobi \textit{thetanull} function \cite{Ramanujan:1998b}, 
\bea
\begin{split}
\theta_3 (x) = \sum_{n=-\infty}^\infty e^{-\pi x n^2}.
\end{split}
\eea
This choice of regularization has the advantage that the first term of (\ref{eq:44}) corresponds to the fully resummed result and the effect of the cut-off is encoded in the latter two terms of (\ref{eq:44}). Proving the consistency of the two approaches, with and without the cut-off, requires care since in the limit $\ell_c \to 0$ {the} theta function diverges and requires regularization. The theta function can be regularized by requiring that the cut-off dependent contribution in (\ref{eq:31er}) vanishes in the limit of $\ell_c \to 0$, corresponding to a subtraction of the divergent contribution. To compute the finite $\ell_c \to 0$ limit of (\ref{eq:31er}) we use the modular transformation 
\bea
\begin{split}
\theta_3\left(x\right) = \sqrt{ 1 / x}~ \theta_3\left({1/ x} \right)
\end{split}
\eea
along with the small $x$ expansion of the theta function \cite{Ramanujan:1998b}, leading to
\bea
\begin{split}
\theta_3\left(x \right) \approx  {1/\sqrt{x}} + O(\exp (-\pi/x)/\sqrt{x}). 
\end{split}
\eea
Using this expression in (\ref{eq:44}) and removing the divergent part, consistently with the regularization of the theta function, gives the expected fully resummed result. The corrections due to the cut-off near $R \sim R_{crit}$ to the fully resummed result can be estimated assuming that $\ell_c \ll R_{crit}$ and can be computed including higher order corrections in the expansion of the theta function:
\bea
\theta_3\left(x \right) -{1/ \sqrt{x}} \approx  {2\sigma / \sqrt{x}} +{2\sigma^4 / \sqrt{x}} + O\left({\sigma^9/ \sqrt{x}}\right),
\label{smallxtheta32}
\eea
where $\sigma = \exp\left(-\pi/x\right)$. Using (\ref{smallxtheta32}) in (\ref{eq:44}) implies that corrections to the fully resummed result are exponentially small, that is the behavior of the vacuum energy is robust against the inclusion of a cut-off smaller than the critical radius for large enough ring size. For small $R$, we expect the cut-off to regulate the diverging $1/R$ behavior of leading term in Eq.~(\ref{eq:31er}). Expanding the theta functions in (\ref{eq:44}) for $\ell_c / R^2$ large, gives at leading order 
\bea
\begin{split}
\tilde{\mathscr{E}_r} \approx {\rho^2 - \eta^{-2} \over 2} e^{-{\ell_c/ \eta^2} } \xrightarrow[R \to 0]{} 0, 
\end{split}
\eea
which can be contrasted with the $\rho^2 \sim R^{-2}$ behavior of the vacuum energy as obtained by full resummation.

\section{Conclusions} 

The behavior of the quantum vacuum energy of an interacting non-relativistic system is far from trivial. Here we have looked at an example of this using a nonlinear Schr\"odinger quantum field theory and computed the quantum vacuum energy and force without resorting to any perturbative expansion in the coupling constant, simply relying on the exact integrability of the nonlinear problem. The novel results are summarized in the ``phase diagram'' of Fig.~\ref{fig:hmap} which shows how the fluctuation-induced force as a function of rotation and interaction strength separates into a noncausal region plus an attractive-repulsive region. This behavior arises from the stabilization between an attractive Casimir-like component and a repulsive centrifugal one.

An interesting potential connection seems evident between our quantum field theoretical set-up and the area of ultracold atoms. A possibly relevant example is the set-up of Ref.~\cite{Eckel:2014}, which consists {a $^{23}$Na BEC confined in a} ring of size $R \sim 20\mu$m. {Considering} a quasi-1D approximation the interaction strength $\lambda$ can be expressed in terms of the scattering length $a_s$ and the transverse length scale {$l$ as $\lambda=g/(\pi l^2)$ (here $g=4\pi\hbar^2 a_s/m$ defines the atomic interaction with $a_s=50 a_0$ for $^{23}$Na) \cite{Samuelis:2000}.} Taking {$l \sim 2\mu$}m to {ensure} that {$l \ll R$} and assuming a condensate of $N=2\times10^3$ atoms (species other than $^{23}$Na, e.g. $^{87}$Rb, can have larger atom numbers and different scattering lengths), allows us to arrive at a dimensionless interaction strength {$\lambda m l/\hbar^2 \sim 4 a_s N / l$}, which, using the above parameter values, gives {$4a_s N/l \sim 10$}, a value close to that used in Fig.~(\ref{fig:hmap}-a). {The force $F_s$ can also be estimated in a similar manner, using Eq.~\eqref{eq:37} and the above definitions we obtain a dimensionless force 
\bea
\begin{split}
F_sml^3/\hbar^2=-(3a_s Nl/8\pi R^2)+m^2l^3 R\Omega^2/4\hbar^2. 
\end{split}
\eea
Using a rotation speed of $\Omega{\sim} 2\pi{\times} 25$Hz from the experiment of Ref.~\cite{Pandey:2019}, we obtain $F_sml^3/\hbar^2\sim 0.1$, modest but potentially large enough to be observable in a future experiment.} {Using these values a ring of size $R \sim 20 \mu$m would fall in the causal repulsive region of Fig.~\ref{fig:hmap}(a), and $\Omega ml^2/\hbar\sim 0.2$} favoring lower rotation frequencies. {In this work Dirichlet boundary conditions have been used, which could be simulated using a weak-link as realized in the BEC ring experiments of Refs.~\cite{Ryu:2013,Jendrzejewski:2014}}. 

 {The physical system described in this work has potential applications in atomtronics \cite{Amico:2021}, facilitating an additional opportunity to explore the fundamental physics associated with the quantum vacuum. Extensions to systems of fermions \cite{Cai:2022} or with multiply-connected geometries \cite{Bland:2022} offers additional avenues to explore the effects described in this work in uncharted scenarios.}

\section*{ACKNOWLEDGEMENTS}
A.F.'s research was supported by the Japanese Society for the Promotion of Science Grant-in-Aid for Scientific Research (KAKENHI, Grant No. 21K03540). M.E.'s research was supported by the Australian Research Council Centre of Excellence in Future Low-Energy Electronics Technologies (Project No. CE170100039) and funded by the Australian Government, and by the Japan Society of Promotion of Science Grant-in-Aid for Scientific Research (KAKENHI Grant No. JP20K14376). {A.F. and M.E. acknowledge support from the University of Queensland (SMP Accelerator Grant).} 
We would like to thank G. Marmorini for  discussions.

\appendix*

\section*{Derivation of the solutions and of the spectrum}

\subsection*{Solutions}

In this appendix, we will illustrate how to solve the system (9)-(10) that we re-write here for convenience:
\bea
\label{eqsm:1}
&&\frac{1}{2mR^2}\rho^{\prime\prime}+\left(\omega_p-\frac{\alpha^{\prime \, 2}}{2mR^2}-\Omega\alpha^\prime \right)\rho-\frac{\lambda}{2}\rho^3=0 
\\
&&\frac{1}{2mR^2}\alpha^{\prime\prime}\rho+\frac{1}{mR^2}\alpha^\prime\rho^\prime+\Omega\rho^\prime=0.
\label{eqsm:2}
\eea
The second equation can be solved by separation of variables, replacing $\alpha^\prime\left(\varphi\right)=\beta\left(\varphi\right)$,
\begin{equation}
\begin{split}
\beta^\prime\rho+2\beta\rho^\prime+2mR^2\Omega\rho^\prime=0.
\end{split}
\end{equation}
Integrating yields
\begin{equation}
\begin{split}
\int\frac{d\beta}{2\beta+2mR^2\Omega}=-\int\frac{d\rho}{\rho},
\end{split}
\end{equation}
leading to
\begin{equation}
\begin{split}
\label{val_alphap}
\alpha^\prime=\beta=\frac{C}{\rho^2}-mR^2\Omega,
\end{split}
\end{equation}
where ${C}$ is an integration constant.
Substituting (\ref{val_alphap}) allows to express (\ref{eqsm:1}) as
\begin{equation}
\begin{split}
\label{ODE_2}
\rho^{\prime\prime}+F_1\frac{1}{\rho^3}+F_2\rho+F_3\rho^3=0.
\end{split}
\end{equation}
where we have defined
\begin{equation}
\begin{split}
\begin{dcases}
\label{subs_F}
F_1=-C^2 \\
F_2\equiv\epsilon_p=2mR^2\left(\omega_p+\frac{m}{2}R^2\Omega^2\right)\\
F_3=-\lambda mR^2
\end{dcases}
\end{split}
\end{equation}
The first integral can be obtained by multiplying both sides of (\ref{ODE_2}) by $\rho^\prime$ and integrating
\begin{equation}
\begin{split}
\label{ODE_2_a}
\rho^{\prime \, 2}-F_1\frac{1}{\rho^2}+F_2\rho^2+\frac{F_3}{2}\rho^4+H=0,
\end{split}
\end{equation}
where $H$ is an integration constant. Finally, multiplying both sides for $\rho^2$ and changing variables,
\begin{equation}
\begin{split}
\begin{dcases}
\label{change_var1}
s=\rho^2 \\
\frac{s^\prime}{2}=\rho\rho^\prime,
\end{dcases}
\end{split}
\end{equation}
gives
\begin{equation}
\begin{split}
\label{ODE_3}
\left(s^\prime\right)^2=4F_1-4Hs-4F_2s^2-2F_3s^3.
\end{split}
\end{equation}
The above equation corresponds to the differential equation of an undamped quadratic anharmonic oscillator, whose canonic form can be obtained by differentiating with respect to $\varphi$ and dividing by $2s^\prime$,
\begin{equation}
\begin{split}
\label{ODE_4}
s^{\prime\prime}=-2H-4F_2 s-3F_3 s^2.
\end{split}
\end{equation}
Following Ref.~\cite{Lakshmanan:2003} (see also Refs.~\cite{Carr:2000::1,Carr:2000::2,Sacchetti:2020,Cominotti:2014}), we can write Eq.~(\ref{ODE_3})
\bea
\label{ODE_6}
\left(s^\prime\right)^2 &=& d\left(s-\alpha_1\right)\left(s-\alpha_2\right)\left(s-\alpha_3\right)\nonumber \\
&=&-d\alpha_1\alpha_2\alpha_3+d\left(\alpha_1\alpha_2+\alpha_2\alpha_3+\alpha_3\alpha_1\right)s-
\nonumber \\&&
-d\left(\alpha_1+\alpha_2+\alpha_3\right)s^2+ds^3
\label{ODE_roots}
\eea
with $\alpha_1,\alpha_2,\alpha_3$ being roots of RHS polynomial, and
\begin{equation}
\begin{split}
\begin{dcases}
\label{system1}
\alpha_1+\alpha_2+\alpha_3=-2\frac{F_2}{F_3} \\
\alpha_1\alpha_2+\alpha_2\alpha_3+\alpha_3\alpha_1=2\frac{H}{F_3} \\
\alpha_1\alpha_2\alpha_3=2\frac{F_1}{F_3}.
\end{dcases}
\end{split}
\end{equation}
The advantage of expressing Eq.~(\ref{ODE_3}) as Eq.~(\ref{ODE_6}) is that the latter is one of the standard nonlinear ordinary differential equation, whose solutions can be expressed in terms of Jacobi elliptic functions as:
\begin{equation}
\begin{split}
s\left(\varphi\right)=\alpha_3-\left(\alpha_3-\alpha_2\right)\sn^2\left(q\varphi,k\right)
\label{sn_1}
\end{split}
\end{equation}
with
\begin{equation}
\begin{split}
k=\sqrt{\frac{\alpha_3-\alpha_2}{\alpha_3-\alpha_1}},
\end{split}
\label{kappa}
\end{equation}
being the elliptic modulus and
\begin{equation}
\begin{split}
q=\sqrt{\frac{F_3}{2}\left(\alpha_3-\alpha_1\right)}.
\end{split}
\label{qqq}
\end{equation}
Throughout the paper, we adopt for the Jacobian elliptic functions the notation of NIST Digital Library of Mathematical Functions \cite{NIST}, according to which
\begin{subequations}
\begin{empheq}[left=\empheqlbrace]{align}
&\dn^2\left(x,m\right)+m^2 \,\sn^2\left(x,m\right)=1 \label{jacob_id_1}
\\
&E\left(m\right)=\int_0^{\frac{\pi}{2}}\sqrt{1-m^2 \sin^2\left(t\right)}dt \label{jacob_id_2}
\\
&K\left(m\right)=\int_0^{\frac{\pi}{2}}\frac{1}{\sqrt{1-m^2 \sin^2\left(t\right)}}dt;
\label{jacob_id_3}
\end{empheq}
\end{subequations}
In the above expressions, $\dn$ and $\sn$ define, respectively, the Jacobi delta amplitude and elliptic sine, and $K\left(m\right)$ and $E\left(m\right)$ define, respectively, the complete elliptic integrals of the first and second kind.

Using (\ref{change_var1}) and (\ref{sn_1}), the general solution can be expressed as follows
\begin{equation}
\begin{split}
\rho\left(\varphi\right)=\sqrt{\alpha_3-\left(\alpha_3-\alpha_2\right)\sn^2\left(q\varphi,k\right)}.
\end{split}
\label{solrho}
\end{equation}
The solution depends on the physical parameters $m, R,\Omega,\lambda, \omega$ and the integration constants $C,H$ through the roots $\alpha_1,\alpha_2,\alpha_3$ and through $k$, and $q$.

\subsubsection*{Boundary conditions, normalization and spectrum}

Having to deal with the solution (\ref{solrho}) in all its algebraic complexity (the explicit dependence of the solutions on the physical parameters is intricate) can be bypassed by directly imposing the boundary conditions. In this paper, we shall focus on the case of Dirichlet boundary conditions,
\begin{equation}
\begin{split}
\Phi\left(t,0\right)=\Phi\left(t,2\pi R\right)=0,
\end{split}
\end{equation}
which implies (using Eqs.~(6) and (8) in the main text)
\begin{equation}
\begin{split}
\rho\left(0\right)=0.
\end{split}
\end{equation}
Exploiting the fact that 
\begin{equation}
\begin{split}
\sn\left(0,k\right)=0, \,\,\,\,\,\,\,\,\, \forall k\in\mathbb{R}
\end{split}
\end{equation}
we can write
\begin{equation}
\begin{split}
\rho\left(0\right)=\sqrt{\alpha_3-\left(\alpha_3-\alpha_2\right)\sn^2\left(0,k\right)}=0,
\end{split}
\end{equation}
which implies
\bea
\alpha_3=0.
\eea
This simplifies the solution to
\begin{equation}
\begin{split}
\label{sol_rho1}
\rho\left(\varphi\right)=\sqrt{\alpha_2}\,\sn\left(\sqrt{-\frac{F_3}{2}\alpha_1}\varphi,\sqrt{\frac{\alpha_2}{\alpha_1}}\right).
\end{split}
\end{equation}
Notice that the constraint $\alpha_3=0$ also fixes the value of the integration constant $C$: specifying $\alpha_3=0$ in equation (\ref{ODE_roots}) gives
\begin{equation}
\begin{split}
\label{ODE_roots2}
\left(s^\prime\right)^2&=d\left(\alpha_1\alpha_2\right)s-d\left(\alpha_1+\alpha_2\right)s^2+ds^3
\end{split}
\end{equation}
where no constant terms appear. Comparing the above equation with Eq.~(\ref{ODE_3}) implies that the constant term proportional to $F_1$ in Eq.~(\ref{ODE_3}) has to vanish:
\begin{equation}
\begin{split}
F_1\equiv-C^2=0.
\end{split}
\label{Ciszero}
\end{equation}
This simplifies considerably the system of equations (\ref{system1}), which become 
\begin{equation}
\begin{split}
\begin{dcases}
\label{system2}
\alpha_1+\alpha_2=-2\frac{F_2}{F_3} \\
\alpha_1\alpha_2=2\frac{H}{F_3}. \\
\end{dcases}
\end{split}
\end{equation}
Imposing the remaining boundary condition at $\varphi=2\pi$, i.e.,
\begin{equation}
\begin{split}
\label{bc_2p}
\rho\left(2\pi\right)=0  
\end{split}
\end{equation}
implies that
\begin{equation}
\begin{split}
\sn\left(\sqrt{-\frac{F_3}{2}\alpha_1}\varphi,\sqrt{\frac{\alpha_2}{\alpha_1}}\right)=0,
\end{split}
\end{equation}
from which, using the the property of the Jacobi $\sn$ function
\begin{equation}
\begin{split}
\sn\left(2nK\left(m\right)+2i l K\left(1-m\right),m\right)=0 \;\;\;\;\;\;\;\;  \forall n,l \in \mathbb{Z}, \nonumber
\end{split}
\end{equation} 
we arrive at
\begin{equation}
\begin{split}
-\frac{F_3}{2}\alpha_1=\frac{n}{\pi}K\left(\sqrt{\frac{\alpha_2}{\alpha_1}}\right).
\end{split}
\end{equation}
$K\left(x\right)$ represents the elliptic integral of the first kind as defined in (\ref{jacob_id_3}). 

Summarizing, we have
\begin{subequations}
\begin{empheq}[left=\empheqlbrace]{align}
\label{system2_a}
&\alpha_1+\alpha_2=-2\frac{F_2}{F_3} \\
\label{system2_b}
&\alpha_1\alpha_2=2\frac{H}{F_3} \\
\label{system2_c}
&-\frac{F_3}{2}\alpha_1=\frac{n}{\pi}K\left(\sqrt{\frac{\alpha_2}{\alpha_1}}\right)
\end{empheq}
\end{subequations}
that, together with the normalization condition, give us 4 independent equations for 4 variables $(\alpha_1,\alpha_2,H,\omega)$. 
In practice, the above set of equations defines the quantization condition of the eigenvalues $\omega_p$.
It is possible to show that the system admits a unique solution but it is easier to proceed in an alternative and faster way. 
Using the condition $F_1=0$ directly in equation (\ref{ODE_2}) leads to
\begin{equation}
\begin{split}
\label{ODE_prin}
\rho^{\prime\prime}+F_2\rho+F_3\rho^3=0.
\end{split}
\end{equation}
This step allows us to write the solution in a simpler form
\begin{equation}
\begin{split}
\label{ans_rho}
\rho\left(\varphi\right)=A\,\sn\left(q\varphi,k\right).
\end{split}
\end{equation}
Using (\ref{ans_rho}) in Eq.~(\ref{ODE_prin}) gives
\bea
&&2 k^2q^2 \sn^3\left(q \varphi,k\right)-\left(1+k^2\right)q^2\sn\left(q \varphi,k\right)=
\nonumber\\&&
-F_2\,\sn\left(q\varphi,k\right)-F_3A^2\,\sn\left(q\varphi,k\right).
\eea
From which we obtain, by matching the coefficients of the like powers of $\sn\left(q\varphi,k\right)$, the following relations
\begin{subequations}
\begin{empheq}[left=\empheqlbrace]{align}
\label{cond_ode_qk1}
&F_3=-\frac{2k^2q^2}{A^2} \\
\label{cond_ode_qk2}
&F_2=\epsilon_p=\left(1+k^2\right)q^2.
\end{empheq}
\end{subequations}
It is interesting to notice that the quantization condition (\ref{system2_c}) becomes 
\begin{equation}
\begin{split}
q_n=\frac{n}{\pi}K\left(k\right), ~~~ n \in \mathbb{N}.
\end{split}
\end{equation}
(The same quantization conditions would have arisen imposing the boundary conditions directly on the simpler solutions, confirming the validity of the procedure). Solving the remaining conditions (\ref{cond_ode_qk1}),(\ref{cond_ode_qk2}) for $q,k$ we obtain the following relations
\begin{subequations}
\begin{empheq}[left=\empheqlbrace]{align}
\label{def_q_gen}
&q_n^2=\frac{A_n^2F_3+2F_2}{2} \\
\label{def_k_gen}
&k_n^2=-\frac{A_n^2F_3}{A_n^2F_3+2F_2} \\
\label{def_q_quant}
&q_n=\frac{n}{\pi}K\left(k_n\right),
\end{empheq}
\end{subequations}
where we have defined $q \rightarrow q_n$, $k \rightarrow k_n$ and $A \rightarrow A_n$ to make the dependence on $n$ explicit.\\ 
~~~The computation of the ``normalization'' coefficients $A_n$ is carried out using the non-relativistic normalization condition,
\begin{equation}
\begin{split}
\braket{\phi|\phi}=\int_V \,dV \phi^\star\left(x\right)\phi\left(x\right)=1,
\end{split}
\end{equation}
which gives
\begin{equation}
\begin{split}
\label{A_calculation}
1&
=A_n^2 R\int_0^{2 \pi} \sn^2\left(q_n\varphi,k_n\right)d\varphi  
\\& =2\pi \frac{A_n^2 R}{k_n^2}-\frac{A_n^2 R}{k_n^2}\int_0^{2 \pi}\dn^2\left(q_n\varphi,k_n\right)d\varphi
\end{split}
\end{equation}
where we have used the Jacobi identity (\ref{jacob_id_1}). Using the definition of the Jacobi epsilon function
\begin{equation}
\begin{split}
\label{iden_4}
&\E\left(x,k\right)=\int_0^x\dn^2\left(t,k\right)\,dt
\end{split}
\end{equation}
and the following relation (which comes from a combination of quasi-addition and quasi-periodic formulas \cite{NIST}), 
\begin{equation}
\begin{split}
\label{iden_5}
&\E\left(n K\left(k\right),k\right)=n E\left(k\right)
\end{split}
\end{equation}
it is possible to simplify the normalization condition as
\begin{equation}
\begin{split}
\label{A_n}
1&=\frac{A_n^2 R}{k_n^2}\left(2\pi-\frac{\pi}{n K\left(k_n\right)}\E\left(2n K\left(k_n\right),k_n\right)\right)\\
&=\frac{2\pi A_n^2 R}{k_n^2}\left(1-\frac{E\left(k_n\right)}{K\left(k_n\right)}
\right)
\end{split}
\end{equation}
or equivalently
\begin{equation}
\begin{split}
\label{norm_coef_nr}
A_n^2=\frac{k_n^2}{2\pi R\left(1-\frac{E\left(k_n\right)}{K\left(k_n\right)}\right)}.
\end{split}
\end{equation}
Using relations (\ref{def_q_gen}), (\ref{def_k_gen}) and (\ref{def_q_quant}), it takes simple steps to arrive at the following relation
\begin{equation}
\begin{split}
\label{fine_1}
-\frac{F_3}{R}\frac{\pi}{4n^2}=K\left(k_n\right)\left(K\left(k_n\right)-E\left(k_n\right)\right)
\end{split}
\end{equation}
that, along with (\ref{cond_ode_qk2}), closes the quantization condition for $\omega_n$, $k_n$ and $q_n$.

\subsubsection*{Complete solution}
In order to find the complete solution, 
\begin{equation}
\begin{split}
f_p\left(\varphi\right)=\rho\left(\varphi\right)e^{i\alpha\left(\varphi\right)},
\end{split}
\end{equation}
we shall need to find the phase $\alpha\left(\varphi \right)$. Using (\ref{val_alphap}) and (\ref{Ciszero}) it is easy to arrive at the following expression
\begin{equation}
\begin{split}
\label{val_alpha_def}
\alpha\left(\varphi\right)=-mR^2\Omega\varphi+\Xi
\end{split}
\end{equation}
where $\Xi$ is an integration constant.
Using (\ref{val_alpha_def}), (\ref{ans_rho}), along with Eqs.~(6), and (8) from the main text, we arrive at 
\begin{equation}
\begin{split}
\Phi\left(t,\varphi\right)=A_n\, e^{-i\omega_n t}e^{-i\left(mR^2\Omega\varphi-\Xi\right)}\sn\left(q_n\varphi,k_n\right).
\end{split}
\end{equation}
The quantity $\Xi=\pi/4$ is a phase and the factor $\exp (i \Xi)$ corresponds to a rotation of the phase $\alpha$, leaving the EOM unaltered.

\subsubsection*{Noninteracting limit}

The noninteracting limit $\lambda \to 0$ simplifies Eq.~(14) from the main text to 
\begin{equation}
\begin{split}
K\left(k_n\right)\left(K\left(k_n\right)-E\left(k_n\right)\right)=0.
\label{eq:17}
\end{split}
\end{equation}
Using the fact that $K\left(m\right)>0$ for any $m\in\left[0,1\right)$, the above condition further simplifies to  $K\left(m\right)=E\left(m\right)$, hence $m=0$, i.e. $k_n \to 0$ for $\lambda \to 0$. Setting $k_n=0$ in the equation for the eigenvalues and using the properties $\sn\left(x,0\right)=\sin\left(x\right)$ and $K\left(0\right)=\pi/2$ yields
\bea
\omega_n=\frac{n^2}{4}\frac{1}{2mR^2}-\frac{m}{2}R^2\Omega^2.
\label{eq:18}
\eea
Taking similar steps in the eigenfunctions, leads to
\bea
\Phi\left(t,\varphi\right)=A_n\, e^{-i\omega_n t}e^{-i\left(mR^2\Omega\varphi-\Xi\right)}\sin\left(\frac{n}{2}\varphi\right),
\label{eq:19}
\eea
which reduce to ordinary plane waves for $\Omega \to 0$. For comparison, see Refs.~\cite{Chernodub:2012em,Schaden:2012,Corradini:2021yha}.

\subsubsection*{Solutions in the laboratory frame}

The solutions in the stationary-laboratory frame can be obtained by performing the inverse coordinate transformation: $t \to t$ and $\varphi \to \varphi + \Omega t$, 
leading to
\bea
\Phi\left(t,\varphi\right) &=& A_n\, e^{-i\omega_n t}e^{-i\left(mR^2\Omega\left[\varphi + \Omega t\right]_{2\pi}-\Xi\right)} \sn\left(q_n\left[\varphi + \Omega t\right]_{2\pi},k_n\right),
\nonumber
\eea
with $\left[ u \right]_{2\pi} \equiv u \left[\mbox{mod($2\pi$)}\right]$, which encodes the $2\pi$ periodicity of the solutions.


\begin{thebibliography}{99}

\bibitem{Takahashi:1969}
Y. Takahashi, H. Shimodaira, 
Nuovo Cimento {\bfseries 62}A (1969) 255


\bibitem{Plunien:1986}
G.~Plunien, B. Muller, W. Greiner,
Phys. Rept. {\bfseries 134} (1986) 87


\bibitem{Ford:1975}
L. H. Ford,
Phys. Rev. D{\bfseries 11} (1975) 3370

\bibitem{Casimir:1948}
H.B.G.~Casimir,
Proc. Kon. Ned. Ak. Wet. {\bfseries 51} (1948) 793

\bibitem{Milton:2001}
K.A. Milton, \emph{The Casimir Effect}, World Scientific (2001)

\bibitem{Bordag:2009}
M.~Bordag, G.L.~Klimchitskaya, U.~Mohideen, V.M.~Mostepanenko,
\emph{Advances in the Casimir Effect}, 
Oxford University Press (2009)

\bibitem{Lamoreaux:1997}
S.K.~Lamoreaux, 
Phys. Rev. Lett. {\bfseries 78} (1997) 5

\bibitem{Mohideen:1998}
U.~Mohideen, A.~Roy,
Phys. Rev. Lett. {\bfseries 81} (1998) 4549


 \bibitem{Gong:2021}
T. Gong, M.R. Corrado, A.R. Mahbub, C. Shelden and J.N. Munday,
{{Nanophotonics} {\bfseries 10(1)} (2021) 523} 

\bibitem{Toms:2012}
D.J. Toms, 
\emph{The Schwinger Action Principle and Effective Action}, 
{Cambridge University Press} (2012)

\bibitem{Toms:2002}
D.J. Toms, 
 {{Phys. Rev. A}{\bfseries 66} (2002) 013619}

\bibitem{Nakayama:2023}
K. Nakayama. K. Suzuki,
Phys. Rev. Research {\bfseries 5} (2023) L022054

\bibitem{Cougo-Pinto:2002}
M.V. Cougo-Pinto, C. Farina, J.F.M. Mendes, A.C. Tort, Braz. J. Phys. {\bfseries 31} (2001) 45

\bibitem{Kolomeisky:2013zra}
E.~B.~Kolomeisky, H.~Zaidi, L.~Langsjoen and J.~P.~Straley,
Phys. Rev. A \textbf{87} (2013), 042519


\bibitem{Corradini:2021yha}
O.~Corradini, A.~Flachi, G.~Marmorini, M.~Muratori and V.~Vitagliano,
J. Phys. A \textbf{54}, no.40, 405401 (2021)

\bibitem{Recati:2005}
A. Recati, J.N. Fuchs, C.S. Pe\c{c}a, W. Zwerger,
Phys. Rev. A\textbf{72} (2005) 023616

\bibitem{Roberts:2005}
D. C. Roberts and Y. Pomeau,
Phys. Rev. Lett. \textbf{95} (2005) 145303

\bibitem{Edery:2006}
A. Edery,
J. Stat. Mech., \textbf{118} (2006) P06007

\bibitem{Schiefele:2009}
J. Schiefele, C. Henkel, 
J. Phys. A \textbf{42}, 045401 (2009)

\bibitem{Biswas:2010}
S. Biswas, J.K. Bhattacharjee, D. Majumder, K. Saha, N. Chakravarty,
J. Phys. B: At. Mol. Opt. Phys. \textbf{43} (2010) 085305 

\bibitem{Schecter:2014}
M. Schecter and A. Kamenev,
Phys. Rev. Lett. \textbf{112} (2014) 155301


\bibitem{Schecter:2015}
A. Diallo, C. Henkel,
J. Phys. B: At. Mol. Opt. Phys. \textbf{48} (2015) 165302.

\bibitem{VanThu:2017}
N. Van Thu, L. Thi Theu,
J. Stat. Mech., \textbf{168} (2017) 1

\bibitem{Marino:2017}
J. Marino, A. Recati, I. Carusotto,
Phys. Rev. Lett. \textbf{118} (2017) 045301

 \bibitem{Song:2021}
P.T.~Song, N.~Van Thu, 
  {{J. Low Temp. Phys.}
  {\bfseries 202} (2021) 160} 

\bibitem{Song:2022}
P.T. Song,
Phys. Lett. A\textbf{455} (2022) 128515

\bibitem{Ford:1979b}
L.~Ford, 
  {{Proc. R. Soc. Lond. A}
  {\bfseries 368} (1979) 305}

\bibitem{Toms:1980a}
D.~J.~Toms, 
  {{Phys.\ Rev.\ D}{\bfseries 21} (1980) 928}

\bibitem{Peterson:1982}
C.~Peterson, T.H.~Hansson, and K.~Johnson, 
  {{Phys. Rev. D}
  {\bfseries 26} (1982) 415}

 \bibitem{Bordag:2021}
M.~Bordag, 
  {{Universe}
  {\bfseries 7(3)} (2021) 55} 

\bibitem{Flachi:2023prd}
A.~Flachi, M.~Edmonds,
{Phys.\ Rev.\ } D{\bf 107} (2023) 025008


\bibitem{Flachi:2013b}
A. Flachi,
Phys.\ Rev.\ Lett.\ {\bf 110} (2013) 060401

\bibitem{Flachi:2017b}
A. Flachi, \textit{et al.},
Phys.\ Rev.\ Lett.\ {\bf 119} (2017) 031601


\bibitem{Flachi:2021b}
A. Flachi and V. Vitagliano,
J. Phys. A: Math. Theor. {\bf 54} (2021) 265401




\bibitem{Chernodub:2012em}
  M.N. Chernodub,
  {Phys.\ Rev.\ } D{\bf 87} (2013) 025021 (2012).

\bibitem{Schaden:2012}
M.~Schaden,
arXiv:1211.2740 [quant-ph].

\bibitem{Ambrus:2014}
E. Ambrus, E. Winstanley,
Phys. Lett. B{\bf 734} (2014) 296


\bibitem{Lakshmanan:2003}
M. Lakshmanan, S. Rajasekar,
\emph{Nonlinear Dynamics: Integrability, Chaos and Patterns}, 
(2003)


\bibitem{Carr:2000::1} L.D. Carr, C.W. Clark, \& W.P. Reinhardt, 
Phys. Rev. A \textbf{62} (2000) 063610

\bibitem{Carr:2000::2} L.D. Carr, C.W. Clark, \& W.P. Reinhardt, 
Phys. Rev. A \textbf{62} (2000) 063611

\bibitem{Sacchetti:2020}
A. Sacchetti, J. Phys. A \textbf{53}, no.38, 385204 (2020)

\bibitem{Cominotti:2014}
M. Cominotti, D. Rossini, M. Rizzi, F. Hekking, \& A. Minguzzi,
Phys. Rev. Lett. \textbf{113} (2014) 025301

\bibitem{odints:book}
E. Elizalde, S.D. Odintsov, A. Romeo, A.A. Bytsenko, S. Zerbini
\emph{Zeta-regularization techniques with applications}, 
World Scientific (1994)



\bibitem{Baltes-Hilf:1976}
H.P.~Baltes, E.R.~Hilf,
\emph{Spectra of Finite Systems},
Bibliographisches Institute (1976)  


\bibitem{Flachi:2008}
See formula (43) of A.~Flachi, T.~Tanaka, 
Phys. Rev. {\bfseries D}78 (2008) 064011

\bibitem{Elizalde:cs_rep}
E.~Elizalde, 
Comm. Math. Phys. {\bfseries 198} (1998) 83


\bibitem{Landau:1980}
L. D. Landau and E. M. Lifshitz, \emph{Fisica Statistica}, Editori Riuniti (1978)


\bibitem{Ramanujan:1998b}
B.C. Berndt, 
\emph{Ramanujan’s Notebooks, Part III}, 
Springer-Verlag, 1991;
\emph{Ramanujan’s Notebooks, Part V}, 
Springer-Verlag, 1998


\bibitem{Eckel:2014}
S. Eckel, \textit{et al.}, 
Nature 506 (2014) 200 

\bibitem{Samuelis:2000}
C. Samuelis, \textit{et al.},
Phys. Rev. A 63, (2000) 012710

\bibitem{Pandey:2019}
S. Pandey, \textit{et al.},
Nature 570, (2019) 205

\bibitem{Ryu:2013}
C. Ryu, \textit{et al}, 
Phys. Rev. Lett. 111, (2013) 205301 

\bibitem{Jendrzejewski:2014}
F. Jendrzejewski, \textit{et al.},
Phys. Rev. Lett. 113, (2014) 045305

\bibitem{Amico:2021}
L. Amico, \textit{et al.},
AVS Quantum Sci. 3, (2021) 039201

\bibitem{Cai:2022}
Y. Cai, \textit{et al.},
Phys. Rev. Lett. 128, (2022) 150401

\bibitem{Bland:2022}
T. Bland, \textit{et al.},
Phys. Rev. Research 4, (2022) 043171




\bibitem{NIST}
NIST Digital Library of Mathematical Functions. https://dlmf.nist.gov/, Release 1.1.9 of 2023-03-15. F. W. J. Olver, A. B. Olde Daalhuis, D. W. Lozier, B. I. Schneider, R. F. Boisvert, C. W. Clark, B. R. Miller, B. V. Saunders, H. S. Cohl, and M. A. McClain, eds.







\end{thebibliography}
\end{document}